# Pedestrian Dead Reckoning System using Quasi-static Magnetic Field Detection

Liqiang Zhang, Kai Guo, Yu Liu

School of Microelectronics, Tianjin University

Tianjin, China

*Abstract*—Kalman filter-based Inertial Navigation System (INS) is a reliable and efficient method to estimate the position of a pedestrian indoors. Classical INS-based methodology which is called IEZ (INS-EKF-ZUPT) makes use of an Extended Kalman Filter (EKF), a Zero velocity UPdaTing (ZUPT) to calculate the position and attitude of a person. However, heading error which is a key factor of the whole Pedestrian Dead Reckoning (PDR) system is unobservable for IEZ-based PDR system. To minimize the error, Electronic Compass (EC) algorithm becomes a valid method. But magnetic disturbance may have a big negative effect on it. In this paper, the Quasi-static Magnetic field Detection (QMD) method is proposed to detect the pure magnetic field and then selects EC algorithm or Heuristic heading Drift Reduction algorithm (HDR) according to the detection result, which implements the complementation of the two methods. Meanwhile, the QMD, EC, and HDR algorithms are integrated into IEZ framework to form a new PDR solution which is named as Advanced IEZ (AIEZ).

## I. INTRODUCTION

Global Navigation Satellite System (GNSS) is widely used as an essential positioning method for pedestrian navigation. But in the field, such as urban canyon or indoor environment, GNSS becomes unreliable due to the signal attenuation, interference and shade. In order to get continuous and reliable positioning results, Local Positioning System (LPS) is being investigated by using various types of methods, such as inertial measurement unit (IMU) based, wireless-based or vision-based technology [1], etc. Among them, IMUs-based PDR methods are preferred due to lower cost, higher reliability, as well as no external infrastructure needed.

Different mounting positions were researched including handheld [2], belt-mounted [3] or foot-mounted [4-8], etc. Unfortunately, low-cost IMU provides short-term stability and the systematic errors increase due to its integration nature. In order to improve the performance of the system, a widely used Kalman-based PDR framework IEZ [4] utilized ZUPT algorithm to estimate velocity errors during stance phase. We also suggested a new stance phase detection algorithm which is named as Dual ZUPT (D-ZUPT) in order to improve the performance of PDR system [8].

For the performance of PDR system, heading is one of the most essential factors which has an impact on its accuracy. However, the heading error is unobservable for IEZ. Utilizing magnetic heading is an effective way to bound the heading drift error. But magnetic disturbances including soft and hard disturbance [9, 10] are stochastic and changing indoors, which are impossible to be predicted. Hence, the Quasi-static Magnetic field Detection (QMD) algorithm was proposed to select pure magnetic field in magnetic disturbance environment and then fuse the data from gyroscope and magnetometer [11]. The method is available to detect magnetic field which is changing slightly and has a good performance. However, it is difficult to recognize the hard magnetic disturbance. There is no denying that it is still an efficient method to improve the heading accuracy of PDR system.

In this paper, QMD algorithm was redesigned in order to solve the problem that the hard magnetic disturbance is difficult to be detected for the classical QMD algorithm in [11]. When the disturbances are detected by QMD, HDR algorithm which only rely on the data from gyroscope is selected to reduce the heading drift error. On the contrary, Electronic Compass (EC) algorithm is a better option to modify the heading error. Referring to [4], following Foxlin's work, to derive a new PDR system, QMD, HDR and EC algorithm are integrated into IEZ to form a novel framework named AIEZ (Advanced IEZ).

## II. METHODOLOGY

This section describes the mechanism of the PDR framework AIEZ. The processing steps are shown as follows. The block diagram of AIEZ is shown in Fig.1 as well as the flow chart in Fig.2.

- The data collected from accelerometer and gyroscope are utilized to detect the stance phase.

- The roll and pitch angle are estimated using the data from accelerometer during the stance phase.

- Meanwhile, when the pure magnetic field is detected by QMD, EC is selected to calculate the optimal heading. On the contrary, HDR is the better option to correct the heading drift error.

- The Extended Kalman Filter (EKF) is updated with velocity and attitude measurement by ZUPT and Attitude Correction (AC) during stance phase, respectively.
- Eventually, we obtain the optimal attitude, velocity and position using inertial navigation mechanism.

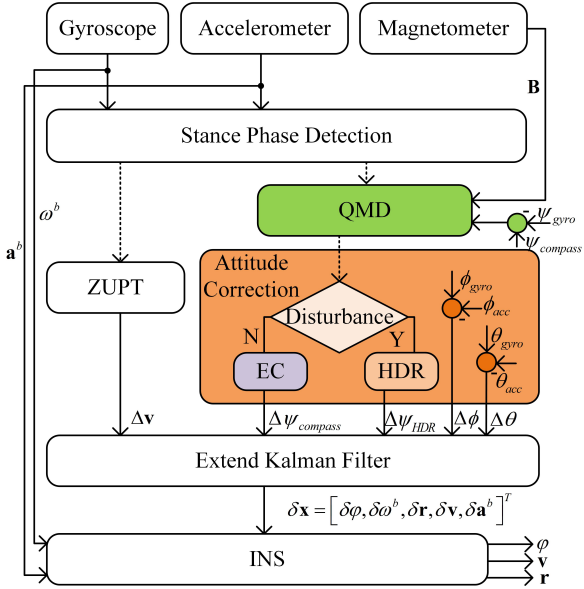

Fig. 1. Block diagram of AIEZ

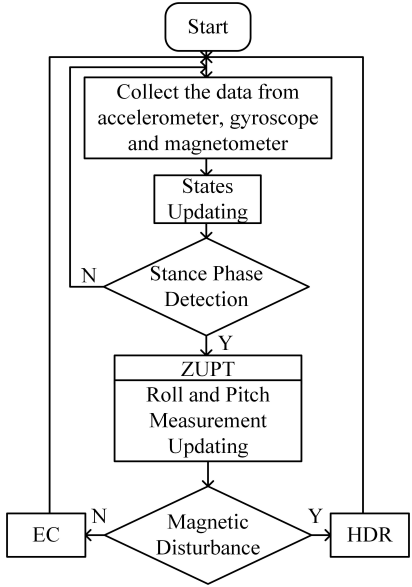

Fig. 2. Flow chart of AIEZ operating

### A. Shoe-mounted Inertial Navigation System

We use standard strap-down Inertial Navigation System (INS) to calculate pedestrian's attitude, velocity and position. The IMU named MTw Awinda with the size of 148×104×31.9 *mm* from Xsens Company is configured to send data at 100 Hz, which is mounted on the heel as shown in Fig.3.

Specific force $\mathbf{a}^b$ and angular rate $\mathbf{\omega}^b$ under body (*b*) frame are utilized to update the attitude, velocity and position of INS. Errors correction is performed by EKF with the 15-element error states $\delta \mathbf{x}_k = [\delta \varphi_k, \delta \omega_k^b, \delta \mathbf{r}_k, \delta \mathbf{v}_k, \delta \mathbf{a}_k^b]^T$, where $\delta \varphi_k$ represents attitude errors, $\delta \omega_k^b$ represents gyroscope bias errors, $\delta \mathbf{r}_k$ represents position errors and $\delta \mathbf{v}_k$ represents velocity, $\delta \mathbf{a}_k^b$ represents accelerometer bias errors as well.

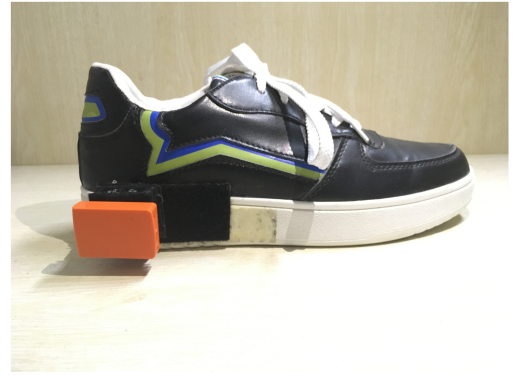

Fig.3. MTw Awinda *(orange color)* and its mounting position

The discretization error state transition model in [6] is

$$\delta \mathbf{x}_{k|k-1} = \mathbf{\Phi}_k \delta \mathbf{x}_{k-1|k-1} + \mathbf{w}_{k-1} \quad (1)$$

where $\delta \mathbf{x}_{k|k-1}$ is the predicted error state, $\delta \mathbf{x}_{k-1|k-1}$ is the last filtered error state, $\mathbf{w}_{k-1}$ is the process noise assuming it is Gaussian white noise.

The state transition matrix is

$$\mathbf{\Phi}_k = \begin{bmatrix} \mathbf{I}_{3\times 3} & \Delta t \mathbf{C}^n_{b_{k|k-1}}(q_{k-1}) & 0 & 0 & 0 \\ 0 & \mathbf{I}_{3\times 3} & 0 & 0 & 0 \\ 0 & 0 & \mathbf{I}_{3\times 3} & \Delta t \mathbf{I}_{3\times 3} & 0 \\ -\Delta t [\hat{\mathbf{a}}_k^n \times] & 0 & 0 & \mathbf{I}_{3\times 3} & \Delta t \mathbf{C}^n_{b_{k|k-1}}(q_{k-1}) \\ 0 & 0 & 0 & 0 & \mathbf{I}_{3\times 3} \end{bmatrix} \quad (2)$$

$$[\hat{\mathbf{a}}_k^n \times] = \begin{bmatrix} 0 & -a_{z,k}^n & a_{y,k}^n \\ a_{z,k}^n & 0 & -a_{x,k}^n \\ -a_{y,k}^n & a_{x,k}^n & 0 \end{bmatrix} \quad (3)$$

where $[\hat{\mathbf{a}}_k^n \times]$ is the skew symmetric matrix for the compensated specific force which has been transformed to the navigation frame. $\Delta t$ is the sample time. $\mathbf{C}^n_{b_{k|k-1}}(q_{k-1})$ represents the predicted rotation matrix that transforms from the *b* frame to the *n* frame, which is updated with quaternion updating.

The discretization quaternion updating model is

$$q_k = \exp\left(\frac{1}{2}\Omega\left(\omega_k^b\right)\Delta t\right)q_{k-1} \qquad (4)$$

where $\Omega\left(\omega_k^b\right)$ is the skew symmetric matrix for the bias-corrected angular rate:

$$\Omega\left(\omega_k^b\right) = \begin{bmatrix} 0 & -\omega_{x,k}^b & -\omega_{y,k}^b & -\omega_{z,k}^b \\ \omega_{x,k}^b & 0 & \omega_{z,k}^b & -\omega_{y,k}^b \\ \omega_{y,k}^b & -\omega_{z,k}^b & 0 & \omega_{x,k}^b \\ \omega_{z,k}^b & \omega_{y,k}^b & -\omega_{x,k}^b & 0 \end{bmatrix} \qquad (5)$$

After every estimation cycle, these error states $\delta\mathbf{x}_k$ derived by EKF compensate the corresponding INS state variables $\mathbf{x}_k$. Compensated specific force is $\hat{\mathbf{a}}_k^b = \mathbf{a}_k^b - \delta\mathbf{a}_k^b$ as well as compensated angular rate $\hat{\omega}_k^b = \omega_k^b - \delta\omega_k^b$.

The filtered attitude is achieved by updating the rotation matrix $\mathbf{C}_{b_{k|k-1}}^n(q_{k-1})$ with the three Euler angle errors $\delta\varphi_k$ estimated by the EKF. Assuming that $\delta\varphi_k$ are small with respect to the $n$ frame, $\mathbf{C}_{b_{k|k}}^n$ can be calculated using *Padé* approximation as follows:

$$\mathbf{C}_{b_{k|k}}^n = \frac{2\mathbf{I}_{3\times 3} + \delta\Theta_k}{2\mathbf{I}_{3\times 3} - \delta\Theta_k}\mathbf{C}_{b_{k|k-1}}^n \qquad (6)$$

where $\delta\Theta_k$ is the skew symmetric matrix for small attitude angles as follows:

$$\delta\Theta_k = \begin{bmatrix} 0 & \delta\varphi_{z,k} & -\delta\varphi_{y,k} \\ -\delta\varphi_{z,k} & 0 & \delta\varphi_{x,k} \\ \delta\varphi_{y,k} & -\delta\varphi_{x,k} & 0 \end{bmatrix} \qquad (7)$$

### B. Stance Phase Detection Algorithm

Stance phase detection is a key process in the PDR framework AIEZ. Better recognition of stance phase contribute to trigger the EKF timely to modify the INS errors for a better performance. Most algorithms for stance phase detection rely on the data from accelerometer [12] or gyroscope [13] as well as both accelerometer and gyroscope [6, 14]. Some scholar implemented the detection using ultrasound, RFID or magnet [15], etc. We utilize the stance hypothesis optimal detection detector (SHOE) proposed in [14] to detect the stance phase.

### C. Quasi-static Magnetic field Detection Algorithm

We formalize the quasi-static magnetic field detection problem as a binary hypothesis testing problem, where the detector can choose between the two hypotheses which are defined as follows:

$$\begin{aligned} H_0 &: \text{Magnetic Disturbance} \\ H_1 &: \text{Pure Magnetic Field} \end{aligned} \qquad (8)$$

Referring to [14], the sensor output model is established as:

$$\mathbf{y}_k = \mathbf{s}_k + \mathbf{v}_k \qquad (9)$$

where $\mathbf{s}_k = \begin{bmatrix} s_k^{\Delta\psi} & s_k^B \end{bmatrix}^T$ and $\mathbf{v}_k = \begin{bmatrix} v_k^{\Delta\psi} & v_k^B \end{bmatrix}^T$. $s_k^{\Delta\psi} \in \mathbb{R}^1$ donates the difference between the heading calculated by INS and the magnetic heading. $s_k^B \in \mathbb{R}^1$ denotes the magnetic field data from magnetometer. Here, $s_k^{\Delta\psi} = |\psi_{k|k-1} - \psi_{compass_k}|$ and $s_k^B = \|\dot{\mathbf{B}}_k\|$. Let $\psi_{k|k-1}$ be the heading which is predicted by INS at time $k$. $\psi_{compass_k}$ represents the magnetic heading which is predicted by EC algorithm at time $k$. $v_k^{\Delta\psi} \in \mathbb{R}^1$ and $v_k^B \in \mathbb{R}^1$ donate the measurement noise associated with $s_k^{\Delta\psi}$ and $s_k^B$, respectively.

Assuming that $\mathbf{v}_k$ is Gaussian white noise, its covariance matrix is:

$$\mathbf{C} = \mathrm{E}\{\mathbf{v}_k \mathbf{v}_k^T\} = \begin{bmatrix} \sigma_{\Delta\psi}^2 & 0 \\ 0 & \sigma_B^2 \end{bmatrix} \qquad (10)$$

where $\sigma_{\Delta\psi}^2$ and $\sigma_B^2$ denote the noise variance of $s_k^{\Delta\psi}$ and $s_k^B$, respectively.

Mathematically, for the two hypotheses, the signal satisfy the conditions that

$$\begin{aligned} H_0 &: \exists k \in \Omega_n \text{ s.t. } s_k^{\Delta\psi} \neq 0 \text{ or } s_k^B \neq 0 \\ H_1 &: \forall k \in \Omega_n \text{ then } s_k^{\Delta\psi} = 0 \text{ and } s_k^B = 0 \end{aligned} \qquad (11)$$

where $\Omega_n = \{l \in \mathbb{N} : n \leq l \leq n + N - 1\}$.

With the sensor model specified by (9), the observations may originate from the family of PDFs (for $i = 0,1$) as follows:

$$p(\mathbf{z}_n; H_i) = \prod_{k \in \Omega_k} p(y_k^{\Delta\psi}; H_i) p(y_k^B; H_i) \qquad (12)$$

where $p(y_k^{\Delta\psi}; H_i) = \frac{1}{\sqrt{2\pi\sigma_{\Delta\psi}^2}} \exp\left(-\frac{1}{2\sigma_{\Delta\psi}^2}(y_k^{\Delta\psi} - s_k^{\Delta\psi})^2\right)$ and $p(y_k^B; H_i) = \frac{1}{\sqrt{2\pi\sigma_{\Delta\psi}^2}} \exp\left(-\frac{1}{2\sigma_{\Delta\psi}^2}(y_k^B - s_k^B)^2\right)$.

Due to the lack of knowledge about the signals $\mathbf{s}_k$, we cannot specify the Probability Distribution Function (PDF) under the two hypotheses completely. Hence, we cannot apply Likelihood Ratio Test (LRT) in *Neyman-Pearson* theorem [16]. By substituting the unknown signal with their Maximum Likelihood Estimates (MLEs) [16], we can derive a Generalized Likelihood Ratio Test (GLRT). The GLRT decides on $H_1$, if

$$L_G(\mathbf{z}_n) = \frac{p(\mathbf{z}_n; H_1)}{p(\mathbf{z}_n; H_0)} > \gamma \qquad (13)$$

Under $H_0$, the signals are completely unknown. In order to maximize $p(\mathbf{z}_n; H_0)$, $\mathbf{s}_k = \{\mathbf{y}_k\}_{k=n}^{n+N-1}$. Thus,

$$p(\mathbf{z}_n; H_0) = \frac{1}{(2\pi\sigma_{\Delta\psi}^2)^{N/2}} \cdot \frac{1}{(2\pi\sigma_B^2)^{N/2}} \quad (14)$$

Under $H_1$, the signals are known, i.e. $\mathbf{s}_k = \begin{bmatrix} 0 & 0 \end{bmatrix}^T$. Thus,

$$p(\mathbf{z}_n; H_1) = \frac{1}{(2\pi\sigma_{\Delta\psi}^2)^{N/2}} \cdot \exp\left(-\frac{1}{2\sigma_{\Delta\psi}^2} \sum_{k \in \Omega_k} (y_k^{\Delta\psi})^2\right) \\ \cdot \frac{1}{\sqrt{2\pi\sigma_B^2}} \exp\left(-\frac{1}{2\sigma_B^2} \sum_{k \in \Omega_k} (y_k^B)^2\right) \quad (15)$$

Hence, combining (13), (14), (15), the GLRT becomes: decide on $H_1$ if

$$L_G(\mathbf{z}_n) = \exp\left(-\frac{1}{2\sigma_{\Delta\psi}^2} \sum_{k \in \Omega_k} (y_k^{\Delta\psi})^2 - \frac{1}{2\sigma_B^2} \sum_{k \in \Omega_k} (y_k^B)^2\right) > \gamma \quad (16)$$

Let $T(\mathbf{z}_n) = -\frac{2}{N} \ln L_G(\mathbf{z}_n)$, we can then state the GLRT as:

$$T(z_n) = \frac{1}{N} \sum_{k \in \Omega_n} \left(\frac{1}{\sigma_{\Delta\psi}^2} (y_k^{\Delta\psi})^2 + \frac{1}{\sigma_B^2} (y_k^B)^2\right) < \gamma' \quad (17)$$

where $\gamma' = -\frac{2}{N} \ln \gamma$. Then,

$$QMD = \begin{cases} 1 & \text{satisfy (12)} \\ 0 & \text{otherwise} \end{cases} \quad (18)$$

### D. Complementary Correction Algorithm

Due to the influence of the biases of accelerometer and gyroscope, the errors of Euler angle accumulate over time. Using the data from accelerometer and gyroscope can calculate the absolute roll and pitch angle when the foot is stationary. Pure magnetic field contributes to a true heading. We use the data during stance phase to modify the attitude calculated by INS.

*1) Roll and Pitch Errors Correction*

During the stance phase, the projection of specific force $\mathbf{a}_k^b$ under $n$ frame should be gravity. Hence, the formula can be described as

$$\begin{bmatrix} a_{x,k}^b \\ a_{y,k}^b \\ a_{z,k}^b \end{bmatrix} = \mathbf{C}_{n_{k|k-1}}^b \begin{bmatrix} 0 \\ 0 \\ g \end{bmatrix} = g \begin{bmatrix} -\sin\theta_{acc,k} \\ \sin\phi_{acc,k}\cos\theta_{acc,k} \\ \cos\phi_{acc,k}\cos\theta_{acc,k} \end{bmatrix}. \quad (19)$$

Then, the roll $\phi_{acc,k}$ and pitch $\theta_{acc,k}$ can be calculated by the data from accelerometer.

$$\phi_{acc,k} = \arctan\frac{a_{y,k}^b}{a_{z,k}^b} \quad (20)$$

$$\theta_{acc,k} = -\arctan\left(\frac{a_{x,k}^b}{\sqrt{(a_{y,k}^b)^2 + (a_{z,k}^b)^2}}\right). \quad (21)$$

Then the roll and pitch error measurements for the EKF are

$$\begin{cases} \Delta\phi_k = \phi_{k|k-1} - \phi_{acc,k} \\ \Delta\theta_k = \theta_{k|k-1} - \theta_{acc,k} \end{cases} \quad (22)$$

where $\phi_{k|k-1}$ and $\theta_{k|k-1}$ denote the predicted roll and pitch by INS, respectively.

*2) Electronics Compass*

The data from magnetometer calibrated by using the software *Magnetic Field Mapper* offered by Xsens are utilized to calculate the magnetic heading. Firstly, the output $\mathbf{B}^b$ of magnetometer under $b$ frame must be transformed to the output $\mathbf{B}^n$ under $n$ frame.

$$\mathbf{B}^n = \begin{bmatrix} \cos\theta_{acc} & \sin\phi_{acc}\sin\theta_{acc} & \cos\phi_{acc}\sin\theta_{acc} \\ 0 & \cos\phi_{acc} & -\sin\phi_{acc} \\ -\sin\theta_{acc} & \sin\phi_{acc}\cos\theta_{acc} & \cos\phi_{acc}\cos\theta_{acc} \end{bmatrix} \mathbf{B}^b \quad (23)$$

Therefore, the horizontal vectors under n frame can be calculated by

$$\begin{cases} B_N^n = B_x^b \cos\theta_{acc} + B_y^b \sin\theta_{acc}\sin\phi_{acc} + B_z^b \cos\phi_{acc}\sin\theta_{acc} \\ B_E^n = B_y^b \cos\phi_{acc} - B_z^b \sin\phi_{acc} \end{cases} \quad (24)$$

Finally, magnetic heading can be calculated using

$$\psi_{compass} = \arctan\left(\frac{B_E^n}{B_N^n}\right) + \psi_d \quad (25)$$

where $\psi_d$ is the magnetic declination at a given point on the earth.

Hence, the heading error measurement for the EKF is

$$\Delta\psi_{compass_k} = \psi_{k|k-1} - \psi_{compass_k} \quad (26)$$

where $\psi_{k|k-1}$ denotes the predicted heading by INS.

*3) Heuristic heading Drift Reduction*

It utilizes the fact that many corridors indoors are straight and people walk along straight line in order to save time and energy. Hence, the HDR algorithm is to detect when a person is walking straight, and in that case apply a correction to the heading error. A simple mechanism is used to recognize the near straight path by

$$\Delta\psi_k = \psi_k - \frac{1}{n}\sum_{s=1}^{n} \psi_{k-s} \quad (27)$$

If $|\Delta\psi_k| < \xi$, we assume that the user is walking along a straight path. Then, HDR is applied to modify the heading error. We make use of $\Delta\psi_{HDR_k}$ as the heading error measurement for the EKF. The formula is described as follows.

$$\Delta\psi_{HDR_k} = \begin{cases} \Delta\psi_k & |\Delta\psi_k| \leq \xi, \text{ straight path} \\ 0 & \text{otherwise, curve path} \end{cases} \quad (28)$$

### E. Extended Kalman Filter

The discretization error state equation is described in section II.A. The processing steps of EKF are shown in Fig.4. Integrating ZUPT [4] (i.e. $\Delta\mathbf{v}_k = \mathbf{v}_{k|k-1} - \begin{bmatrix} 0 & 0 & 0 \end{bmatrix}^T$, where $\mathbf{v}_{k|k-1}$ denotes the velocity which is prior to the EKF updating under $n$ frame at time $k$) and Complementary

Correction algorithm, an EKF-based framework which is named as AIEZ is implemented. The measurement matrix would be a 6×15 matrix:

$$\mathbf{H} = \begin{bmatrix} \mathbf{I}_{3\times3} & \mathbf{0}_{3\times3} & \mathbf{0}_{3\times3} & \mathbf{0}_{3\times3} & \mathbf{0}_{3\times3} \\ \mathbf{0}_{3\times3} & \mathbf{0}_{3\times3} & \mathbf{0}_{3\times3} & \mathbf{I}_{3\times3} & \mathbf{0}_{3\times3} \end{bmatrix} \quad (29)$$

The error measurement would be:

$$\mathbf{z}_k = [\Delta\varphi_k, \Delta\mathbf{v}_k]^T \quad (30)$$

where the attitude error measurement is

$$\Delta\varphi_k = [\Delta\phi_k, \Delta\theta_k, QMD\cdot\Delta\psi_{compass_k} + (1-QMD)\Delta\psi_{HDR_k}]^T \quad (31)$$

and the velocity error measurement is $\Delta\mathbf{v}_k$.

The covariance matrix of measurement noise is

$$\mathbf{R}_k = diag(\sigma_{roll}^2, \sigma_{pitch}^2, (QMD\cdot\sigma_{compass}^2 + (1-QMD)\cdot\sigma_{HDR}^2), \boldsymbol{\sigma}_v^2) \quad (32)$$

where diag() denotes a diagonal matrix, $\sigma_{roll}^2$ denotes the variance of roll error updating, $\sigma_{pitch}^2$ denotes the variance of pitch error updating, $\sigma_{compass}^2$ denotes the variance of magnetic heading error calculated by EC algorithm, $\sigma_{HDR}^2$ denotes the variance of the heading error calculated by HDR algorithm and $\boldsymbol{\sigma}_v^2$ denotes the variance of the velocity error measured by stance phase detection algorithm.

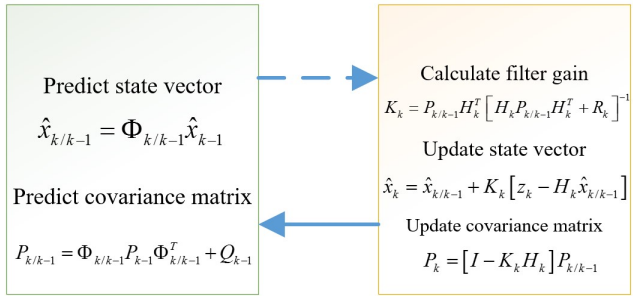

Fig. 4. Processing steps of EKF

## III. EXPERIMENTAL RESULTS

The system consists of a consumer-grade IMU mounted on the heel of the shoe (as shown in Fig.3) with the transmission speed at 100 Hz via wireless mode to the laptop for logging and real-time visualization. The IMU named Mtw Awinda contains 3-axis MEMS accelerometer, 3-axis MEMS gyroscope and 3-axis MEMS magnetometer. The parameters in detail can be found in the website of Xsens company. Several experiments were performed to access the performance of the proposed methodology.

### A. Walking in the Hard Magnetic Disturbance Field

In this case, a pedestrian walks along the corridor with the magnet placed on the floor as shown in Fig.5. The red circles in Fig.6 indicates that the hard magnetic disturbance happens due to the wrong magnetic heading while the INS heading keeps stable. Owing to the characteristic of the magnet, the magnetic field intensity near it is constant. Accordingly, from the zoomed drawing in Fig.6, the magnetic headings stay stable but wrong. The proposed QMD algorithm contributes to a right detection results compared with the classical QMD method and has a better performance.

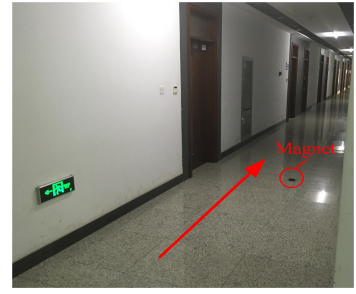

Fig. 5. Walking in the Hard Magnetic Disturbance Field

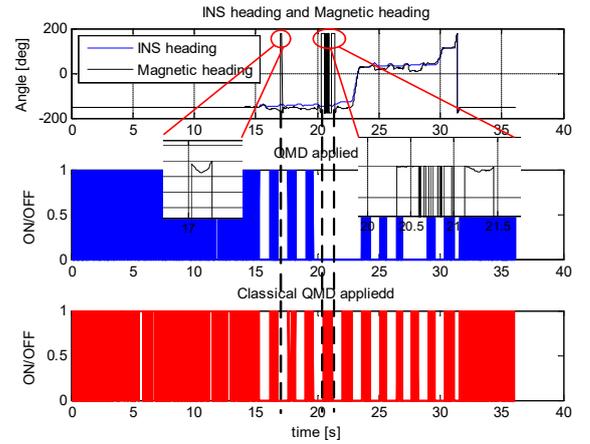

Fig. 6. Comparision of the detection results of classical QMD and proposed method

### B. Walking in Real Indoor Environment

In this case, we walked along the regular trajectory for three cycles for 7 minutes in the teaching building 26 in Tianjin University. The total distance of the first trajectory is about 515.5 meters. We evaluate the performance among the IEZ, IEZ+classical QMD and AIEZ (as shown in Fig.7) and logged the Total Travelled Distance (TTD) error in Table I.

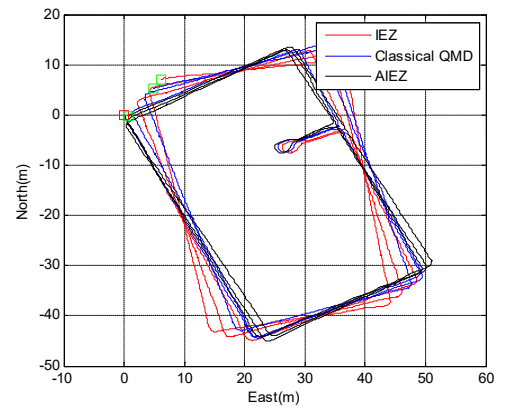

Fig. 7. The trajectory of IEZ, IEZ+classical QMD and AIEZ

TABLE I.    POSITIONING ERRORS OF DIFFERENT METHODS

|  | Position Error (m) | TTD error (%) |
|---|---|---|
| IEZ | 9.49 | 1.84 |
| IEZ+classical QMD | 7.09 | 1.38 |
| AIEZ | 0.79 | 0.15 |

IV. CONCLUSION AND FUTURE WORKS

In this paper, QMD was implemented to combine the EC algorithm and the HDR algorithm in order to achieve complementation between them. Classical QMD algorithm fails to recognize the hard magnetic field and mistakes it for pure magnetic field. Our proposed QMD method captures the moment when the magnetic field intensity is stable and the magnetic heading is similar to the INS heading. Hence, it have a better performance than classical QMD. AIEZ framework fuses the EC and HDR algorithm and is superior to IEZ and IEZ+classical QMD frameworks. In the future, plenty of experiments need to be settled to evaluate the performance of QMD detector and AIEZ framework.